\documentclass[twocolumn]{jpsj3} 
\usepackage{subfigure}
\usepackage{widetext}
\usepackage{bm}
\usepackage{color}

\newcommand{\HH}{\mathcal{H}}
\newcommand{\bT}{\textbf{T}}
\newcommand{\bt}{\textbf{t}}
\newcommand{\kk}{\textbf{k}}
\newcommand{\Q}{\textbf{Q}}
\newcommand{\be}{\begin{eqnarray}}
\newcommand{\ee}{\end{eqnarray}}
\newcommand{\onlinecite}[1]{\hspace{-1 ex} \nocite{#1}\citenum{#1}} 
\title{Symmetry and Gap Classification of the non-symmorphic SrPtAs}

\author{\textsc{Mark H Fischer}$^{1}$\thanks{E-mail address: mark.h.fischer@gmail.com} and \textsc{Jun Goryo}$^{2}$}

\inst{$^{1}$Department of Condensed Matter Physics, The Weizmann Institute of Science, Rehovot 760001, Israel \\
$^{2}$Department of Advanced Physics, Hirosaki University, Hirosaki 036-8561, Japan}

\abst{%
The hexagonal superconductor SrPtAs exhibits time-reversal-symmetry breaking below $T_c$, hinting at an unconventional pairing state.
Therefore, the symmetry of the underlying crystal is important for the classification of possible gap structures and their mixing.
Here, we use the generating point group D$_{6h}$ of SrPtAs for a comprehensive classification of the gap functions and to construct a tight-binding model.
Our work clarifies questions of symmetry and topology in this non-symmorphic material and allows for better comparison with other hexagonal systems.
}

\kword{SrPtAs, hexagonal symmetry, chiral d-wave superconductivity, non-centrosymmetric superconductivity}

\begin{document}
\maketitle

\section{Introduction}
The pnictide superconductor SrPtAs~\cite{nishikubo:2011} ($T_c=2.4K$) has attracted attention recently due to the time-reversal-symmetry (TRS) breaking occurring with the onset of superconductivity.\cite{youn:2012, goryo:2012, biswas:2013,fischer:2014a,matano:2014, akbari:2014, wang:2014, bruckner:2014,tutuncu:2014}
Unlike other pnictide superconductors, SrPtAs has a hexagonal crystal structure. 
This has important consequences for the unconventional superconducting order parameters which  can be classified with respect to the irreducible representations of the hexagonal point group and are only allowed to mix within the same representation.\cite{sigrist:1991} 
A particularly interesting example is the degeneracy of $d_{xy}$ and $d_{x^2-y^2}$ which allows for a TRS-breaking ($d+id$)-wave superconducting state.\cite{fischer:2014a}

SrPtAs crystallizes in the $P6_3/mmc$ space group (\#194). This space group is non-symmorphic, i.e., some point group operations have to be combined with non-trivial translations to map the crystal onto itself, with a generating point group isomorphic to $D_{6h}$. 
Note that considering a unit cell containing two Pt-As layers with an inversion center in between leads to a point group $D_{3d}\subset D_{6h}$.~\cite{goryo:2012} 
However, focusing solely on this unit cell and its point group misses the symmetry operations in $D_{6h}\setminus D_{3d}$ and with it half the irreducible representations. 
The full symmetry also has to be considered to construct a (tight-binding) Hamiltonian, which is responsible for the gap mixing on a microscopic level.

In this article, we apply a comprehensive symmetry analysis of the superconducting order parameters in {SrPtAs} using the generating point group $D_{6h}$. 
In section \ref{sec:sym} we first elaborate on the symmetry of SrPtAs and discuss a tight-binding model for illustration and to introduce basis functions.
In section \ref{sec:op}, we use the symmetry to classify gap functions and analyze their intermixing. 
Finally, we discuss and summarize the resulting gap functions in section \ref{sec:discussion}.
 
\begin{figure}[b]
  \centering
  \subfigure[]{
  \includegraphics[width=0.25\textwidth]{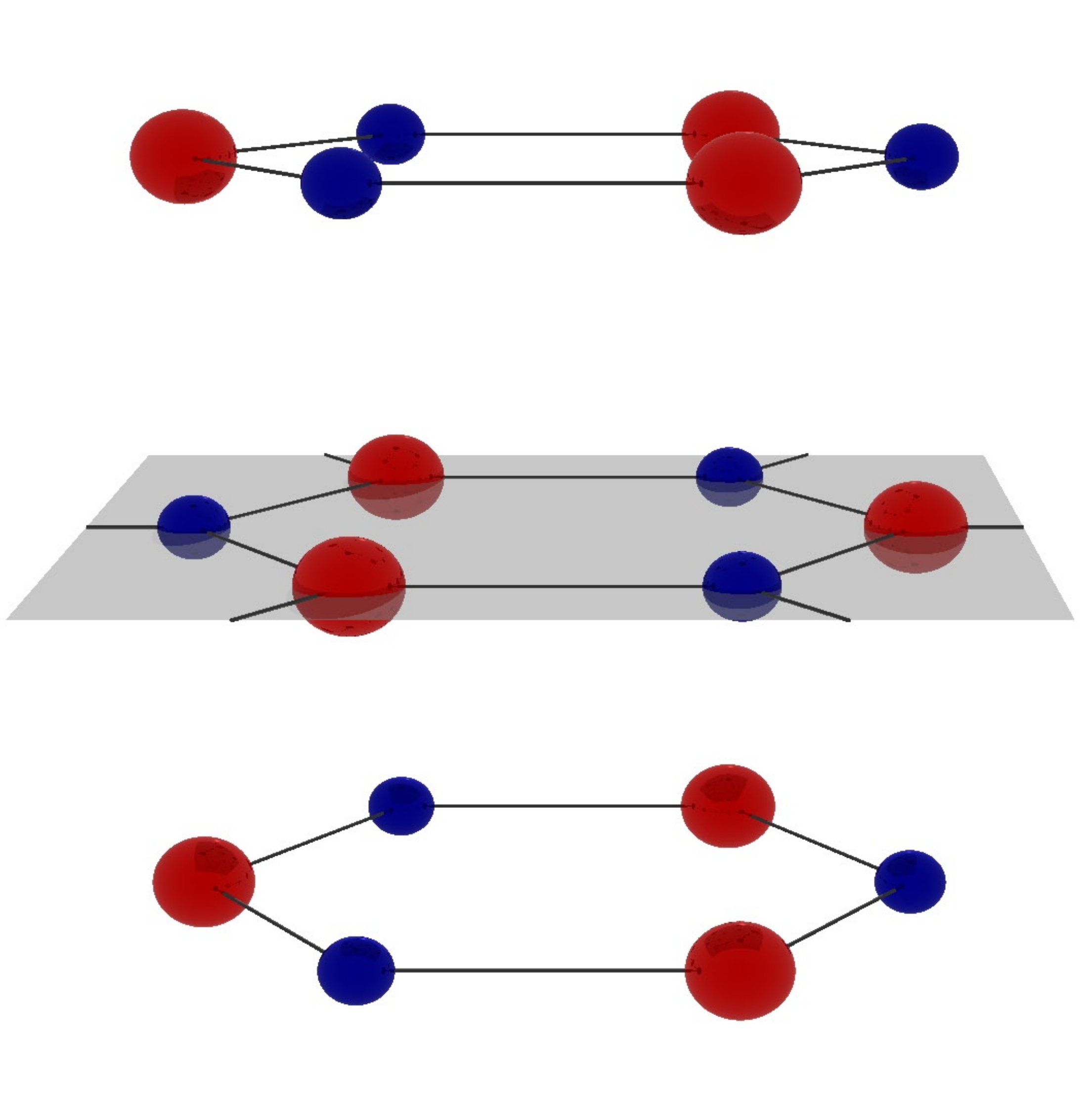}
}
 \subfigure[]{
   \includegraphics[width=0.2\textwidth]{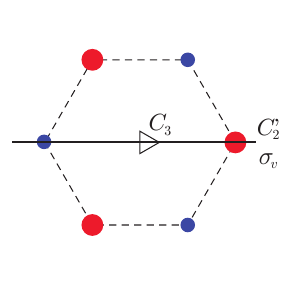}
}
  \caption{(a) 3D crystal structure only showing Pt (red) and As (blue) sites. (b) Symmetry of a single layer with a three-fold rotation axis (triangle) and 3 two-fold rotation axes and 3 mirror planes (only one of each shown).}
  \label{fig:lattice}
\end{figure}

\section{Symmetry and tight-binding Hamiltonian}
\label{sec:sym}
SrPtAs possesses a hexagonal structure composed of honeycomb Pt-As layers, see Fig.~\ref{fig:lattice}(a), which are spaced by Sr layers.
In a single honeycomb layer, the A and B sites are occupied by Pt and As, respectively, such that there is no center of inversion.
However, in the neighboring layers, the Pt and As sites are interchanged, and the alternating stacking of the layers results in global centers of inversion between each two neighboring layers. 
Due to this stacking, the crystal has a non-symmorphic space group. 
An important consequence of this is that it is not possible to choose a unit cell that possesses the full generating point group $D_{6h}$ of the system. 

To illustrate this property and better understand the resulting spin-orbit coupling and its effects, we here follow Ref.~\onlinecite{fischer:2011b} and choose as a starting point a single Pt-As layer, Fig.~\ref{fig:lattice}(b).  The point group of such a single layer contains the symmetry operations $\{E, 2C_3, 3C_2', \sigma_h, 2 S_3, 3\sigma_v\}=D_{3h}$, in particular it lacks inversion $i$. The full crystal, however, has a generating point group that is isomorphic to $D_{\rm 6h}$. While the symmetry transformations of $D_{\rm 3h}$ with respect to the Pt-As layer leave the full crystal invariant, the elements of $D_{\rm 6h}\setminus D_{\rm 3h}=\{2C_6, C_2, 3C_2'', i, 2S_6, 3\sigma_d\}$ interchange the two distinct layers and thus have to be combined with a translation along the $z$ axis by $\tilde{c} = c/2$.

\begin{figure}[tb]
  \centering
  \includegraphics[width=0.25\textwidth]{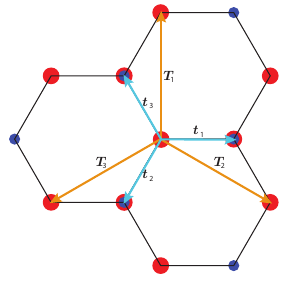}
  \caption{Lattice vectors with $\bT_n$ connecting Pt sites within the same layer and $\bt_n$ the in-plane vectors connected to nearest-layer hopping.}
  \label{fig:vectors}
\end{figure}

Following this construction, we discuss a tight-binding Hamiltonian with `$s$' orbitals on the Pt sites.~\cite{youn:2012, youn:2012b}
Starting from a single layer (with point group $D_{3h}$), the Hamiltonian contains a hopping term on the triangular (Pt) lattice,
\begin{equation}
  \HH = t \sum_{\kk s} \Big[\sum_{n}\cos(\bT_n\cdot\kk)\Big]c^{\dag}_{\kk s}c^{\phantom{\dag}}_{\kk s},
  \label{eq:Hsingle}
\end{equation}
where we have introduced the lattice vectors
\begin{eqnarray}
  \bT_1 &=& a(0,1,0),\\
  \bT_2 &=& a(\sqrt{3}/2, -1/2, 0),\\
  \bT_3 &=& a(-\sqrt{3}/2, -1/2, 0)
  \label{eq:Ts}
\end{eqnarray}
[see Figure \ref{fig:vectors}] and $c^\dag_{\kk s}$ creates an electron with momentum $\kk$ and spin $s$. In addition, there is a spin-orbit-coupling (SOC) term due to the As positions,
\begin{equation}
  \HH^{\rm soc} = \alpha_{\rm so}\sum_{\kk, s, s'}\Big[\sum_{n}\sin(\bT_n\cdot\kk)\Big]c^{\dag}_{\kk s}\sigma^3_{ss'}c^{\phantom{\dag}}_{\kk s'}
  \label{eq:HSOC}
\end{equation}
with $\sigma^i$ the Pauli matrices acting in spin space.
Note that from a symmetry perspective, the hopping term~\eqref{eq:Hsingle} transforms as $A_{1g}$ in $D_{6h}$, while the spin-orbit term~\eqref{eq:HSOC} transforms as $B_{1u}$. In $D_{\rm 3h}$, however, both transform as A$_{1g}$ and are thus symmetry allowed here.

To treat the layer staggering of the 3D structure, we divide SrPtAs into two sublattices, namely even and odd layers, respectively.~\cite{fischer:2011b} Instead of working in layer space, we fold the Brillouin zone (BZ) in $z$ direction (with respect to stacked layers without staggering) by introducing the operators 
\begin{equation}
  c_{\alpha\kk s} = \left\{\begin{array}{ll} c_{\kk s} & \alpha=1 \\ c_{\kk+\Q s}&\alpha=2\end{array}\right.,
  \label{eq:caks}
\end{equation}
where $\Q=(0,0,\pi)$ (setting $\tilde{c}=1$), and Pauli matrices ($\tau$) acting in $\{\kk, \kk+\Q\}$ space (see Appendix).
The spin-independent Hamiltonian can then be written as
\begin{equation}
  \HH = \sum_{\alpha,\alpha'}\sum_{\kk, s}\vphantom{\sum}'\HH_{\kk\alpha\alpha'}c^{\dag}_{\alpha\kk s}c^{\phantom{\dag}}_{\alpha'\kk s},
  \label{eq:H0}
\end{equation}
where the sum $\sum_{\kk}'$ runs over the folded BZ. $\HH_{\kk\alpha\alpha'}$ consists of a trivial intra-sublattice hopping
\begin{equation}
  \HH_{\kk}^{\rm intra} = [t\sum_{n}\cos(\bT_n\cdot\kk) + t_z' \cos(2k_z)]\tau^0,
  \label{eq:Hintra}
\end{equation}
and an inter-sublattice hopping connecting neighboring layers,
\begin{equation}
  \HH_{\kk}^{\rm inter} = t_z\cos k_z [\sum_{n}\cos(\bt_n\cdot\kk)\tau^3 + \sum_{n}\sin(\bt_n\cdot\kk)\tau^2].
  \label{eq:Hinter}
\end{equation}
Note the momentum structure due to the fact that Pt sites of neighboring layers do not sit on top of each other [see Fig.~\ref{fig:vectors}].
Finally, the spin-orbit-coupling term has opposite sign on the two sublattices, reading in this basis
\begin{equation}
  \HH_{\kk}^{\rm soc} = \alpha_{\rm so} \sum_{n}\sin(\bT_n\cdot\kk) \sigma^3\otimes\tau^1.
  \label{eq:Hsoc}
\end{equation}
The nature of the $\tau$ matrices and their transformation behavior is summarized in Table~\ref{tab:taus} and a detailed derivation of the tight-binding terms can be found in the appendix. 
Note that all the bands resulting from this Hamiltonian are doubly degenerate and $S_z$ is conserved. Other spin-orbit-coupling terms that break $S_z$ are connected to inter-layer hopping and small, and we will discuss consequences of them later.

In terms of symmetry, the Hamiltonian has to transform trivially, i.e., as a scalar, under all the symmetry transformations of the crystal (corresponding to $A_{1g}$ in $D_{6h}$). This is achieved here through the combination of a momentum and spin part with a $\tau$ matrix such that the whole Hamiltonian transforms as $A_{1g}$, i.e. either $A_{1g}\otimes A_{1g}$  [see Eq.~\eqref{eq:Hintra}] or $B_{1u}\otimes B_{1u}$ [Eq.~\eqref{eq:Hsoc}].\cite{footnote:dorbitals} In addition, note that by construction a function with $f(\kk) = f(\kk+\Q)$ [$f(\kk) = - f(\kk+\Q)$] has to be combined with $\tau^0$ or $\tau^1$ ($\tau^2$ or $\tau^3$), respectively.

\begin{table}[tb]
  \centering
  \begin{tabular}{cccc}
    \hline\hline
                & intra-sublattice & inter-sublattice & IR \\
    \hline
    intra-band   &  $\tau^0$       &      $\tau^3$   & $A_{1g}$\\
    inter-band   &  $\tau^1$       &      $\tau^2$   & $B_{1u}$\\
  \end{tabular}
  \caption{Classification of the Pauli matrices $\tau^a$ defined in the space $\{\kk, \kk+\Q\}$ of the two bands in the folded Brillouin zone. The sublattice refers to even and odd layers.}
  \label{tab:taus}
\end{table}

\section{Gap Classification}
\label{sec:op}
A general gap function in the $\{\kk, \kk+\Q\}$ basis of Eq.~\eqref{eq:caks} can be written as
\begin{equation}
  \Delta_{ss'}^{\alpha\alpha'}(\kk) = \{\psi_a(\kk)(i\sigma^y) + [\vec{d}_a(\kk)\cdot\vec{\sigma}](i\sigma^y)\}_{ss'}\otimes\tau^a_{\alpha\alpha'},
  \label{ea:generalgap}
\end{equation}
and the spin-singlet part $\psi_a(\kk)$ and the spin-triplet part $\vec{d}_a(\kk)$ can be classified according to the irreducible representations of $D_{6h}$. Note that due to the SOC, we have to classify the combined momentum and spin part for the spin-triplet gap functions. 
For the full classification, we use the same scheme as for the Hamiltonian, namely that gaps transform as $R\otimes R'$ with $R'$ the irreducible representation $\tau^a$ transforms as, i.e. either $A_{1g}$ or $B_{1u}$ (Table~\ref{tab:taus}).
However, there are two requirements for the gap functions $\psi_a(\kk)$ and $\vec{d}_a(\kk)$: (1) As noted in the previous section, our construction requires $\psi_a(\kk)$ and $\vec{d}_a(\kk)$ for $a=2,3$ to change sign under $\kk\mapsto\kk+\Q$ but not for $a=0,1$. (2) The Pauli principle has to be satisfied, requiring that for an even spin-singlet and odd spin-triplet function $a=0,1,3$ (``triplet in $\{\kk, \kk+\Q\}$''), and $a=2$ for odd spin-singlet and even spin-triplet order parameters (``singlet in $\{\kk, \kk+\Q\}$'').  

A list of possible (tight-binding) functions for the spin-singlet functions $\psi_{a}(\kk)$ respecting the crystals symmetry for the various irreducible representations of $D_{6h}$ is given in Table~\ref{tab:bfpsi}. Table~\ref{tab:bfdz} lists spin-triplet gap functions with a $d$ vector in the $\hat{z}$ direction. Finally, Table~\ref{tab:bfdxy} lists spin-triplet order parameters with in-plane $d$ vector. Note that this corresponds to the classification of Ref.~\onlinecite{sigrist:1991} when looking at their expansion in $k_x, k_y, k_z$, e.g.,
\begin{eqnarray}
  \sum_n \omega_n\sin(\bT_n\cdot\kk) &\sim& (k_x - i k_y),\\
\sum_n \omega_n\cos(\bT_n\cdot\kk)&\sim& [(k_x^2-k_y^2) + 2 i k_xk_y]
 \label{eq:expansion}
\end{eqnarray}
($\omega_n \!=\! e^{i 2\pi n/3}$), and similarly for the other combinations.
\begin{table}
  \centering
  \begin{tabular}{c|c}
    $\Gamma^+$ & $\psi_{0,1,3}(\kk)$\\
    \hline
    $A_{1g}$ & $1$, $\cos k_z \sum_n\cos(\bt_n\cdot\kk)$, $\sum_n\cos(\bT_n\cdot\kk)$\\
    $A_{2g}$ & - \\ 
    $B_{1g}$ & - \\ 
    $B_{2g}$ & $\sin k_z \sum_n\sin(\bt_n\cdot\kk)$ \\ 
    $E_{1g}$ & $\{\sin k_z\sum_n\omega_n\sin(\bt_n\cdot\kk), \sin k_z \sum_n\omega_n^*\sin(\bt_n\cdot\kk)\}$ \\ 
             & $\{\sin 2 k_z\sum_n\omega_n\sin(\bT_n\cdot\kk), \sin 2k_z \sum_n\omega_n^*\sin(\bT_n\cdot\kk)\}$ \\ 
    $E_{2g}$ & $\{\cos k_z\sum_n \omega_n\cos(\bt_n\cdot\kk),\cos k_z\sum_n \omega_n^*\cos(\bt_n\cdot\kk)\}$ \\ 
             & $\{\sum_n \omega_n\cos(\bT_n\cdot\kk),\sum_n \omega_n^*\cos(\bT_n\cdot\kk)\}$ \\ 
    \hline
    $\Gamma^-$ & $\psi_{2}(\kk)$ \\
    \hline
    $A_{1u}$ & - \\ 
    $A_{2u}$ & $\sin k_z\sum_n\cos(\bt_n\cdot\kk)$\\ 
    $B_{1u}$ & $\cos k_z\sum_n\sin(\bt_n\cdot\kk)$\\ 
    $B_{2u}$ & - \\ 
    $E_{1u}$ & $\{\cos k_z \sum_n\omega_n\sin(\bt_n\cdot\kk), \cos k_z\sum_n\omega_n^*\sin(\bt_n\cdot\kk)\}$ \\ 
    $E_{2u}$ & $\{\sin k_z\sum_n\omega_n\cos(\bt_n\cdot\kk), \sin k_z \sum_n\omega_n^*\cos(\bt_n\cdot\kk)\}$ \\ 
  \end{tabular}
  \caption{Basis functions for $\psi_a(\kk)$ in $D_{6h}$ with $\omega_n = \exp(i 2\pi n/3)$. Note that the table only respects the Pauli principle. In addition, it is required that $\psi_3(\kk)=-\psi_3(\kk+\Q)$. }
  \label{tab:bfpsi}
\end{table}

\begin{table}
  \centering
  \begin{tabular}{c|c}
    $\Gamma^+$ & $\vec{d}_{2}(\kk)$\\
    \hline
    $A_{1g}$ & - \\
    $A_{2g}$ & $\hat{z}\cos k_z \sum_n\cos(\bt_n\cdot\kk)$\\ 
    $B_{1g}$ & - \\ 
    $B_{2g}$ & - \\ 
    $E_{1g}$ & $\{\hat{z}\sin k_z\sum_n\omega_n\sin(\bt_n\cdot\kk), \hat{z}\sin k_z \sum_n\omega_n^*\sin(\bt_n\cdot\kk)\}$ \\ 
    $E_{2g}$ & $\{\hat{z}\cos k_z \sum_n \omega_n\cos(\bt_n\cdot\kk),\hat{z}\cos k_z \sum_n \omega_n^*\cos(\bt_n\cdot\kk)\}$ \\ 
    \hline
    $\Gamma^-$ & $\vec{d}_{0,1,3}(\kk)$ \\
    \hline
    $A_{1u}$ & $\hat{z}\sin k_z\sum_n\cos(\bt_n\cdot\kk)$, $\hat{z}\sin 2k_z$ \\ 
    $A_{2u}$ & - \\ 
    $B_{1u}$ & $\hat{z}\sum_n\sin(\bT_n\cdot\kk)$\\ 
    $B_{2u}$ & $\hat{z}\cos k_z \sum_n\sin(\bt_n\cdot\kk)$\\ 
    $E_{1u}$ & $\{\hat{z}\cos k_z \sum_n\omega_n\sin(\bt_n\cdot\kk), \hat{z}\cos k_z\sum_n\omega_n^*\sin(\bt_n\cdot\kk)\}$ \\ 
    & $\{\hat{z}\sum_n\omega_n\sin(\bT_n\cdot\kk), \hat{z} \sum_n\omega_n^*\sin(\bT_n\cdot\kk)\}$ \\ 
    $E_{2u}$ & $\{\hat{z}\sin k_z\sum_n\omega_n\cos(\bt_n\cdot\kk), \hat{z}\sin k_z \sum_n\omega_n^*\cos(\bt_n\cdot\kk)\}$ \\ 
  \end{tabular}
  \caption{Basis functions for $d^z_a(\kk)$ in $D_{6h}$. Note that the table only respects the Pauli principle. In addition, it is required that $\vec{d}_3(\kk)=-\vec{d}_3(\kk+\Q)$.}
  \label{tab:bfdz}
\end{table}

\begin{table}
  \centering
  \begin{tabular}{c|c}
    $\Gamma^+$ & $\vec{d}_{2}(\kk)$\\
    \hline
    $A_{1g}$ & $\sin k_z[\hat{x}_{+}\sum_n\omega_n\sin(\bt_n\cdot\kk) - \hat{x}_{-}\sum_n\omega_n^*\sin(\bt_n\cdot\kk)]$ \\ 
    $A_{2g}$ & $\sin k_z [\hat{x}_{+}\sum_n\omega_n\sin(\bt_n\cdot\kk) +\hat{x}_{-}\sum_n\omega_n^*\sin(\bt_n\cdot\kk)]$ \\ 
    $B_{1g}$ & $\cos k_z[\hat{x}_{+}\sum_n\omega_n\cos(\bt_n\cdot\kk) + \hat{x}_{-}\sum_n\omega_n^*\cos(\bt_n\cdot\kk)]$ \\ 
    $B_{2g}$ & $\cos k_z[\hat{x}_{+}\sum_n\omega_n\cos(\bt_n\cdot\kk) - \hat{x}_{-}\sum_n\omega_n^*\cos(\bt_n\cdot\kk)]$ \\ 
    $E_{1g}$ & $\{\hat{x}\cos k_z\sum_n\cos(\bt_n\cdot\kk), \hat{y}\cos k_z \sum_n\cos(\bt_n\cdot\kk)\}$ \\ 
    $E_{2g}$ & $\{\sin k_z\hat{x}_{-}\sum_n\omega_n\sin(\bt_n\cdot\kk), \sin k_z\hat{x}_{+}\sum_n\omega_n^*\sin(\bt_n\cdot\kk)\}$ \\ 
    \hline
    $\Gamma^-$ & $\vec{d}_{0,1,3}(\kk)$ \\
    \hline
    $A_{1u}$ & $\hat{x}_{+}\sum_n\omega_n\sin(\bT_n\cdot\kk) -\hat{x}_{-}\sum_n\omega_n^*\sin(\bT_n\cdot\kk)$ \\ 
    $A_{2u}$ & $\hat{x}_{+}\sum_n\omega_n\sin(\bT_n\cdot\kk) + \hat{x}_{-}\sum_n\omega_n^*\sin(\bT_n\cdot\kk)$ \\ 
    $B_{1u}$ & $\sin k_z[\hat{x}_{+}\sum_n\omega_n\cos(\bt_n\cdot\kk) - \hat{x}_{-}\sum_n\omega_n^*\cos(\bt_n\cdot\kk)]$ \\ 
    $B_{2u}$ & $\sin k_z[\hat{x}_{+}\sum_n\omega_n\cos(\bt_n\cdot\kk) + \hat{x}_{-}\sum_n\omega_n^*\cos(\bt_n\cdot\kk)]$ \\ 
    $E_{1u}$ & $\{\hat{x}_{-}\sin 2 k_z, \hat{x}_{+}\sin 2 k_z\}$ \\ 
             & $\{\hat{x}_{-}\sin k_z \sum_n\cos(\bt_n\cdot\kk), \hat{x}_{+}\sin k_z\sum_n\cos(\bt_n\cdot\kk)\}$ \\ 
	     $E_{2u}$ & $\{\hat{x}_{-}\sum_n\omega_n\sin(\bT_n\cdot\kk), \hat{x}_{+}\sum_n\omega^*_n\sin(\bT_n\cdot\kk)\}$ \\ 
  \end{tabular}
  \caption{Basis functions for $\vec{d}_a(\kk)$ in $D_{6h}$. For simplicity we have introduced $\hat{x}_{\pm} = (\hat{x}\pm i\hat{y})$. Note that the table only respects the Pauli principle. In addition, it is required that $\vec{d}_3(\kk)=-\vec{d}_3(\kk+\Q)$.}
  \label{tab:bfdxy}
\end{table}

Utilizing Tables~\ref{tab:bfpsi}-\ref{tab:bfdxy}, we can now construct order parameters belonging to any given irreducible representation of $D_{\rm 6h}$. In the following, we look at some examples for illustration, namely $A_{1g}$, $B_{1u}$, and $E_{2g}$, that have been discussed for SrPtAs~\cite{goryo:2012, biswas:2013, akbari:2014, fischer:2014a, matano:2014, wang:2014,bruckner:2014} starting with $A_{1g}$: assuming intra-layer coupling, the gap function has a dominant term of the form
\begin{equation}
  \Delta^{(0)}_{A_{1g}}(\kk) = \psi_{0}^{A_{1g}}(\kk)(i\sigma^y)\otimes\tau^0
   \label{eq:A1g}
\end{equation}
with 
\begin{equation}
  \psi^{A_{1g}}_{0}(\kk)= A + B \sum_n\cos(\bT_n\cdot\kk).
  \label{eq:psiA1g0}
\end{equation}
$A$ and $B$ are coefficients depending on the pairing interaction and bandstructure.
Order parameters with the same symmetry can be mixed in by the Hamiltonian, i.e., 
\begin{multline}
    \Delta'_{A_{1g}}(\kk) = \tilde{\psi}_{0}^{A_{1g}}(\kk)(i\sigma^y)\otimes\tau^0+ [\vec{d}{\,}^{B_{1u}}_{1}(\kk)\cdot\vec{\sigma}](i\sigma^y)\otimes\tau^1 \\ + (i\sigma^y)\otimes\Big[\psi_{3}^{A_{1g}}(\kk)\tau^3+ \psi^{B_{1u}}_2(\kk)\tau^2\Big].
  \label{eq:A1gb}
\end{multline}
These include an additional intra-sublattice spin-singlet
\begin{equation}
   \tilde{\psi}^{A_{1g}}_{0}(\kk) \propto \cos 2k_z
\end{equation}
and an intra-layer spin-triplet gap function
\begin{equation}
  \vec{d}{\,}^{B_{1u}}_{1}(\kk) \propto \hat{z}\sum_n\sin(\bT_n\cdot\kk), 
  \label{eq:dB1u1}
\end{equation}
as well as a trivial ($A_{1g}$) and a non-trivial ($B_{1u}$) inter-layer spin-singlet
\begin{equation}
  \psi^{A_{1g}}_{3}(\kk) = C \cos k_z \sum_n\cos(\bt_n\cdot\kk)
  \label{eq:psiA1g3}
\end{equation}
and
\begin{equation}
  \psi_2^{B_{1u}}(\kk) = C  \cos k_z \sum_n\sin(\bt_n\cdot\kk).
  \label{eq:psibB1u}
\end{equation}
These latter two gap functions need to be mixed equally in order to respect translational symmetry of the lattice.  
Note that as a result of the intermixing of singlet and triplet gap functions, the resulting order parameter is non-unitary, i.e., $\Delta \Delta^\dag \not\propto\sigma^0\otimes\tau^0$.

In a similar way, the $f$-wave gap function of $B_{1u}$ symmetry has a dominant intra-layer part
\begin{equation}
  \Delta^{(0)}_{B_{1u}}(\kk) = [\vec{d}{\,}^{B_{1u}}_{0}(\kk)\cdot\vec{\sigma}](i\sigma^y)\otimes \tau^0
\end{equation}
with
\begin{equation}
  \vec{d}{\,}^{B_{1u}}_{0}(\kk) \propto \hat{z}\sum_n\sin(\bT_n\cdot\kk). 
  \label{eq:fB1u1}
\end{equation}
The additionally allowed term is
\begin{equation}
  \Delta'_{B_{1u}}(\kk) =  \psi^{A_{1g}}_{1}(\kk)(i\sigma^y)\otimes\tau^1,  \label{eq:b1u}
\end{equation}
an intra-sublattice spin-singlet gap function with
\begin{equation}
  \psi^{A_{1g}}_{1}(\kk)= A + B \sum_n\cos(\bT_n\cdot\kk) + C\cos 2k_z.
  \label{eq:psiA1g1}
\end{equation}

Finally, we look at the ($d\pm id$)-wave order parameter transforming as $E_{2g}$. The gap function reads
\begin{equation}
  \Delta^{(0)}_{E_{2g},\pm}(\kk) = \psi^{E_{2g}}_{0, \pm}(\kk)(i\sigma^{y})\otimes\tau^0
  \label{eq:psiE2g}
\end{equation}
with
\begin{equation}
  \psi^{E_{2g}}_{0,+}(\kk) \propto \sum_n\omega_n \cos (\bT_n\cdot \kk) 
  \label{eq:psiE2g0}
\end{equation}
and $\Delta_{E_{2g},-}(\kk) = [\Delta_{E_{2g},+}(\kk)]^*$.
We can additionally mix in 
\begin{multline}
    \Delta'_{E_{2g},\pm}(\kk) = (i\sigma^{y})\otimes\Big[\psi^{E_{2g}}_{3, \pm}(\kk)\tau^3+ \psi^{E_{1u}}_{\pm}(\kk)\tau^2\Big]\\ + [\vec{d}{\,}^{E_{1u}}_{\pm}(\kk)\cdot\vec{\sigma}](i\sigma^y)\otimes\tau^1,
  \label{eq:psiE2gb}
\end{multline}
including an inter-layer $(d\pm id)$ spin-singlet gap function
\begin{equation}
  \psi^{E_{2g}}_{3,+}(\kk) = A \cos k_z \sum_n\omega_n\cos(\bt_n\cdot\kk),
  \label{eq:psiE2g3}
\end{equation}
as well as $p\pm ip$ gap functions with spin-singlet
\begin{equation}
  \psi^{E_{1u}}_{+}(\kk) = A \cos k_z \sum_n\omega_n\sin(\bt_n\cdot\kk)
  \label{eq:psiE1u}
\end{equation}
and spin-triplet form
\begin{equation}
  \vec{d}{\,}^{E_{1u}}_{+} \propto \hat{z} \sum_n \omega_n \sin(\bT_n\cdot \kk).
  \label{eq:dE1u}
\end{equation}
Again, the two inter-layer gap functions have to be added equally due to translational symmetry constraints.

Before we finish by discussing the above gap functions we note that the conservation of $S_z$ led to a block-diagonal form of the mean-field Hamiltonian. However, this conservation is not imposed by symmetry. Indeed, we can add an additional SOC term to the Hamiltonian due to next-nearest-layer hopping that has both an $S_z$-conserving term,
\begin{equation}
    \HH_{\rm soc}' = \alpha_{\rm so}' \cos 2 k_z \sum_n \sin (\bT_n \cdot k) \sigma^3 \otimes \tau^1,
    \label{eq:fullsoc1}
\end{equation}
as well as an $S_z$-breaking term
\begin{multline}
    \HH_{\rm soc}'' = i\alpha_{\rm so}'' \sin 2 k_z\Big[ (\sigma^1+i\sigma^2) \sum_n \omega_{n} \cos (\bT_n \cdot k) \\ 
    - (\sigma^1-i\sigma^2) \sum_n \omega_{n}^* \cos (\bT_n \cdot k)\Big] \otimes \tau^1.
    \label{eq:fullsoc2}
\end{multline}
This term also allows us to additionally mix in spin-triplet order parameters with in-plane $d$ vectors from Table~\ref{tab:bfdxy}.
The $B_{1u}$ gap function can have an additional inter-layer contribution
\begin{widetext}
\begin{multline}
    \Delta_{B_{1u}}^{\rm inter}(\kk) \propto \sin k_z\Big\{(\sigma^x + i \sigma^y)(i\sigma^y)\otimes\Big[\sum_n\omega_n\sin(\bt_n\cdot\kk)\tau^2 + \sum_n\omega_n\cos(\bt_n\cdot\kk)\tau^3\Big] \\ 
	- (\sigma^x - i \sigma^y)(i\sigma^y)\otimes\Big[\sum_n\omega_n^*\sin(\bt_n\cdot\kk)\tau^2 + \sum_n\omega_n^*\cos(\bt_n\cdot\kk)\tau^3\Big].
  \label{eq:dA1g}
\end{multline}
\end{widetext}
Similarly, we can add the inter-sublattice $E_{1u}$ gap functions of Table~\ref{tab:bfdxy} with $\tau^1$ into the $d\pm id$ order parameter.

\section{Discussion and Summary}
\label{sec:discussion}
We finish our analysis by discussing two important properties of the order parameters above, namely their nodal structure and their topological classification~\cite{schnyder:2008,kitaev:2009,ryu:2010}. For this purpose, we first only consider the dominant in-plane contributions given in Eqs.~\eqref{eq:psiA1g0},  \eqref{eq:fB1u1}, and \eqref{eq:psiE2g0}. In a second step, we will discuss consequences of additional intermixing. As the $A_{1g}$ order parameter is trivial,  we only discuss the $E_{2g}$ and the $B_{1u}$ gap functions.

We first consider the $B_{1u}$ order parameter. A gap of pure $f$-wave symmetry [$\vec{d}_0^{B_{1u}}(\kk)$] has line nodes on any pocket around the $\Gamma$-$A$ line. However, admixture of the other components, e.g. the $\vec{d}_2^{A_{1g}}(\kk)$ and the $\psi_1^{A_{1g}}(\kk)$ components, can lift these nodes. Triplet superconductors in general do not have symmetry imposed line nodes, which is known as Blunt's theorem.~\cite{blount:1985} Note, however, the difference to the $p$-wave situation on a square lattice: While in the latter cases, two gap functions which are related by lattice symmetries are mixed, the $B_{1u}$ channel requires either a singlet component or an inter-layer gap mixed in. Both are subdominant and hence there will still be a significantly suppressed gap along lines on the Fermi surfaces around $\Gamma$-$A$. From a topological point of view the `pure' $B_{1u}$ gap has a block-diagonal form and belongs to class AIII. Mixing through the spin-orbit coupling conserving $S_z$ leaves this classification invariant, while the full spin-orbit coupling of the form~\eqref{eq:fullsoc} destroys the block-diagonal form and leads to an order parameter of class DIII. Finally, a TRS-breaking combination of the various gap functions, which could be realized at a second transition,~\cite{sigrist:1998} would lead to classes A and D, respectively (see Table~\ref{tab:topology}).

Due to the mixing of $d_{xy}$ and $d_{x^2-y^2}$ basis functions the spin-singlet gap with $E_{2g}$ symmetry only has symmetry-imposed point nodes, where the lines parallel to $z$ through $K$, $K'$, and $\Gamma$ intersect with the Fermi surfaces. The pure $d+id$ gap belongs to class C according to the topological classification. The admixing of a triplet $p$-wave, Eq.~\eqref{eq:dE1u}, then changes the classification to A. This class is trivial in three dimensions, but has a $\mathbb{Z}$ classification in two dimensions. For Fermi surfaces with point nodes, such as is the case for SrPtAs, this state is thus a Weyl superconductor. At the point nodes at the BZ boundary, the low-energy excitations can be described as pairs of Majorana-Weyl fermions with a linear spectrum, that do not mix due to $S_z$ conservation.~\cite{fischer:2014a}. Note, however, that the full spin-orbit-coupling term again destroys the block-diagonal form of the mean-field Hamiltonian, resulting in an order parameter belonging to class D (see Table~\ref{tab:topology}). A direct consequence is that the two linear branches of the Majorana-Weyl spectrum can in principle mix, leading to a more complicated nodal and low-energy structure. The gapless character in the nodal region is, however, still protected, as the two nodes on the same point in momentum space carry an equal topological charge.

To summarize, we have presented a comprehensive symmetry analysis of possible gap functions of SrPtAs. Our results can be used for a thorough analysis of instabilities and response functions such as the spin-susceptibility in this complicated material and can help to determine the intriguing superconducting order parameter in SrPtAs.

\begin{table}[tt]
  \centering
  \begin{tabular}{cccc}
    \hline\hline
      Irr. Rep.   & no SOC & $S_z$ SOC & full SOC\\
    \hline
    $E_{2g}$ & C       &  A & D \\
    $B_{1u}$ & AIII   & AIII (A) & DIII (D) \\
   \end{tabular}   \caption{Topological classification~\cite{schnyder:2008,kitaev:2009,ryu:2010} of the $E_{2g}$ and $B_{1u}$ pairing states 
  in SrPtAs for a mean-field Hamiltonian without spin-orbit coupling, the dominant spin-orbit coupling with $S_z$ conservation, and the full spin-orbit coupling, respectively. The classification in brackets for $B_{1u}$ corresponds to a time-reversal-symmetry-breaking mixing of the various basis functions.}
  \label{tab:topology}
\end{table}

\section*{Acknowledgment}
We would like to thank Yuval Baum, Titus Neupert and Manfred Sigrist for helpful discussions. MHF was supported by the Swiss Society of Friends of the Weizmann Institute of Science.
JG is financially supported by a Grant-in-Aid for Scientific Research from the Japan Society for the Promotion of Science, Grant No. 23540437.
\appendix

\section{Tight-Binding Hamiltonian}

In this appendix, we present some details of the derivation of the non-interacting Hamiltonian Eqs. (\ref{eq:Hintra}), (\ref{eq:Hinter}), and (\ref{eq:Hsoc}). 
Note first that only every second layer looks the same, while neighboring layers have the locations of the Pt and As sites interchanged, see Fig.~\ref{fig:lattice}(a). First, we separate the lattice into two sublattices, namely the layers with layer number $l$ even (e) and odd (o), respectively. 
We consider a Hamiltonian on the Pt sites only with intra-layer NN hopping 
\begin{equation}
    \mathcal{H}^{\rm nn}=\frac{t}{2}\sum_{l,i, n}\sum_{s}\left(c^\dag_{\bm r_i^l,s}c^{\phantom{\dag}}_{\bm r^l_i+\bT_n,s}+h.c.\right)
\end{equation}
and next-nearest-layer hopping
\begin{equation}
    \mathcal{H}^{\rm nnl}=\frac{t_z^\prime}{2}\sum_{i,l,s} \left(c^\dag_{\bm r_i^l,s}c^{\phantom{\dag}}_{\bm r_i^{l +2},s}+h.c.\right),
\end{equation}
where $c^\dag_{\bm r^l_i, s}$ creates an electron at $\bm r_i^l = (x_i, y_i, l)$ on the $l$-th layer. 
Writing the electron operators in momentum space,
\begin{equation}
    c_{\bm r^{2m}_i,s}={\sqrt{\frac2N}}\sum_{\kk}^{\rm }\vphantom{\sum}'e^{i \kk \cdot \bm r^{2m}_i}c_{{\rm e}\kk s}
\end{equation}
(and analogly for $l=2m+1$ and $c_{{\rm o}\kk s}$), where $N$ is the number of Pt sites and the sum $\sum_{\kk}'$ covers a BZ corresponding to a unit cell containing both layers, we find
\begin{eqnarray}
  \HH^{\rm intra} \!\!\!\!\!&=&\!\!\!\!\! \sum_{\kk,s}\vphantom{\sum}'\Big[t \sum_{n}\cos(\bT_n\cdot\kk) + t_z' \cos(2k_z)\Big]\times \nonumber \\ 
   &&\qquad\qquad\qquad \times(c^\dagger_{{\rm e}\kk s} c^{\phantom{\dag}}_{{\rm e}\kk s}+c^\dagger_{{\rm o}\kk s} c^{\phantom{\dag}}_{{\rm o}\kk s})\nonumber\\
   &=& \!\!\!\!\! \sum_{\kk,s}\vphantom{\sum}'\Big[t \sum_{n}\cos(\bT_n\cdot\kk) + t_z' \cos(2k_z)\Big] \times\nonumber \\ 
   &&\qquad\qquad \times (c^\dagger_{\kk s} c^{\phantom{\dag}}_{\kk s}+c^\dagger_{\kk+\Q s} c^{\phantom{\dag}}_{\kk+\Q s})
  \label{eq:intrahop}
\end{eqnarray}
with $c_{\kk /  \kk+\Q s} = (c_{{\rm e}\kk s} \pm c_{{\rm o}\kk s})/\sqrt{2}$ and $\Q=(0,0,\pi)$.

Next, we consider the hopping connecting neighboring layers, which starting from real space yields
\begin{eqnarray}
\mathcal{H}^{\rm inter}\!\!\!\!\!\!&=&\!\!\!\!\!\frac{t_z}{2} \sum_{i,n,l,s}\!\left(c^\dagger_{\bm r_i^l,s} c^{\phantom{\dag}}_{\bm r_i^{l+1}+(-1)^l \bt_n,s}+h.c.\right)
\label{eq:inter}\nonumber\\
&=&\!\!\!\!\! t_z \sum_{\bm k, n, s}\!\!\vphantom{\sum}'\!\cos k_z\Big[\cos (\kk \cdot \bt_n) (c^\dagger_{\kk s} c^{\phantom{\dag}}_{\kk s}-c^\dagger_{\kk+\Q s} c^{\phantom{\dag}}_{\kk+\Q s})
\nonumber\\
&& -i  \sin (\kk \cdot \bt_n) (c^\dagger_{\bm k s} c^{\phantom{\dag}}_{\kk+\Q  s}-c^\dagger_{\kk+\Q s} c^{\phantom{\dag}}_{\kk  s})\Big].
\end{eqnarray}
Finally, the staggered spin-orbit coupling term with $\alpha_{\rm so}^l = (-1)^l\alpha_{\rm so}$ reads 
\begin{equation*}
\mathcal{H}^{\rm soc}=\alpha_{\rm so} \sum_{\kk, n, s,s^\prime}\!\!\!\!\!\vphantom{\sum}'  \sin(\kk \cdot \bT_n)\sigma^3_{ss^\prime} (c^\dagger_{\kk s}  c^{\phantom{\dag}}_{\kk+\Q  s^\prime}+c^\dagger_{\kk+\Q s} c^{\phantom{\dag}}_{\kk  s^\prime}),
\end{equation*}
We thus see that the layer dependence $(-1)^l$ of the hopping terms and the spin-orbit coupling gives rise to the mixing of Bloch states with $\kk$ and $\kk + \Q$. This leads to the Eqs. (\ref{eq:Hintra}), (\ref{eq:Hinter}), and (\ref{eq:Hsoc}) in the folded Brillouin zone containing two layers and the doubling of the electron species as given by Eq.~(\ref{eq:caks}).


\begin{thebibliography}{99}
\bibitem{footnote:dorbitals} A similar construction can be used to include a two-orbital basis consisting of $d_{xy}$ and $d_{x^2-y^2}$ orbitals.
\end{thebibliography}
\end{document}